\documentstyle[times,balanced,epsf,prl,aps]{revtex}

\begin{document}
\draft 
\tightenlines

\title{Noise rectification in quasigeostrophic forced turbulence}

\author{Alberto \'Alvarez$^{+}$, Emilio Hern\'andez-Garc\'\i a$^{*}$, 
 and Joaqu\'\i n Tintor\'e$^{*}$}
\address{$^+$ Department of Physics, National Central University,
Chung-Li, Taiwan 32054, Republic of China \\
$^*$ Instituto Mediterr\'aneo de Estudios Avanzados, IMEDEA\cite{imedea} 
(CSIC-UIB)\\  
 E-07071 Palma de Mallorca, Spain}
\date{January 30, 1998}
\maketitle

\begin{abstract}
We study the appearance of large-scale mean motion sustained by stochastic
forcing on a rotating fluid (in the
quasigeostrophic approximation) flowing over topography. As in other noise
rectification phenomena, the effect requires nonlinearity and absence of 
detailed balance to occur. By application of an analytical  
coarse-graining procedure we identify the physical mechanism producing such
effect: It is a forcing coming from the small scales that manifests in 
a change in the effective viscosity operator and in the effective noise 
statistical properties.  
\end{abstract}

\pacs{PACS numbers: 05.40.+j, 92.90.+x, 47.27.-i}

\begin{twocolumns}

Nonlinear interactions can organize random inputs
of energy into coherent motion. This noise-rectification phenomenon has been 
discussed in several contexts 
ranging from biology to physics or engineering 
\cite{directed}. Three ingredients are needed to obtain this kind of
noise-sustained directed motion: nonlinearity, random noise lacking the 
property
of detailed balance, and some symmetry-breaking feature establishing a  
preferred direction of motion. 

It has been recently shown numerically \cite{alvarez97} that directed
motion sustained by noise also appears in quasigeostrophic two-dimensional 
fluid flow over topography. The average flow at large scales approaches a
state highly correlated with topography that disappears if noise or
nonlinearity are switched-off. The presence of topography provides the
symmetry-breaking ingredient needed to fix a preferred direction.
The small scales of the flow follow a more irregular behavior. 

In this Letter we analytically calculate a closed effective equation
of motion for the large scales of the flow, by coarse-graining the
small scales. From this effective equation, the forcing of the small scales on
the large ones (sustained by noise and mediated by topography) appears clearly 
as the responsible for the directed currents. The effect is more noticeable for
increasing nonlinearity, and appears as a renormalization of the viscosity
operator in such a way that it favors relaxation towards a state correlated 
with topography, instead of towards rest. This forcing vanishes when noise
satisfies detailed balance, as in the other kinds of noise-rectification
phenomena. 

The particular model considered here is the equation describing 
quasigeostrophic forced turbulence. A large amount of rotating fluid problems 
concerning 
planetary atmospheres and oceans involve situations in which vertical 
velocities are small and slaved to the horizontal motion 
\cite{williams78,pedlosky87}. Under these circumstances flow patterns can be
described in terms of two horizontal coordinates, the vertical depth 
of the fluid becoming a dependent variable. Although the fluid displays many 
of the 
unique properties of two-dimensional turbulence, some of the aspects of
three-dimensional dynamics are still essential, leading to a quasi
two-dimensional dynamics. In particular, bottom topography appears explicitly
in the equations. This kind of quasi-twodimensional dynamics is not only of
relevance to the case of rotating neutral fluids but there is also a direct
correspondence with drift-wave turbulence in plasma physics 
\cite{mima78,hasegawa79}. 

The streamfunction $\psi({\bf x},t)$, with ${\bf x} \equiv (x,y)$, in the 
quasigeostrophic approximation is governed by the dynamics: 
\cite{pedlosky87}:  

\begin{equation}
\label{eq}
{\partial \nabla^2 \psi \over \partial t} + 
 \lambda \left[ \psi , \nabla^2 \psi +h \right] =
\nu \nabla^4 \psi + F \ , 
\end{equation}
where $\nu$ is the viscosity parameter, $F({\bf x},t)$ is any kind of 
relative-vorticity external forcing, and  $h=f \Delta H/H_0$, with $f$ the 
Coriolis parameter,  
$H_0$ the mean depth, and $\Delta H({\bf x})$ the
local deviation from the mean depth. $\lambda$ is a bookkeeping parameter 
introduced to allow perturbative expansions in the interaction term. The 
physical case corresponds to $\lambda=1$.  
The Poisson bracket or Jacobian is defined as 
\begin{equation}
\label{jac}
[A,B]= {\partial A \over \partial x} {\partial B \over \partial y} - 
{\partial B \over \partial x}{\partial A \over \partial y}\ . 
\end {equation}
%The streamfunction provides the horizontal
%components of the fluid  velocity $\left( u({\bf x}),v({\bf x}) \right)$ 
%from  
%\begin{equation}
%u = - {\partial \psi \over \partial y}\ , \ \ \ 
%v =  {\partial \psi \over \partial x}
%\end{equation}

Equation (\ref{eq}) represents 
the time evolution of the relative vorticity subjected to forcing and 
dissipation. In the case of drift-wave turbulence for a plasma in a
strong magnetic field
applied in the direction perpendicular to ${\bf x}$, ${\psi}$ is 
related to the electrostatic potential, and $h=\ln( \omega_{c}/n_0)$, 
where $\omega_{c}$ and $ n_0 $ are
the cyclotron frequency and plasma density respectively. Eq. (\ref{eq}) is 
also the limiting
case of the more general Charney-Hasegawa-Mima equation when the scales are 
small compared to the ion
Larmor radius or the barotropic Rossby 
radius\cite{hasegawa79,hasegawa85,horton90}. 

We now stablish how the dynamics of long-wavelength 
modes in (\ref{eq}), when $F$ is a random forcing, is affected by the small 
scales. Stochastic 
forcing has been used in fluid dynamics problems to model stirring
forces\cite{marti97}, wind forcing\cite{battisti}, short scale 
instabilities\cite{williams78}, thermal noise 
\cite{treiber96,ll}, or processes
below the resolution of computer models \cite{mason94}, among others
\cite{mccomb95}. A useful choice of $F$, flexible enough to model a variety of 
processes, is to assume $F$ to be  Gaussian stochastic process with zero
mean and correlations given by 
$\left< \hat F_{\bf k}(\omega)\hat F_{\bf k'}(\omega' ) \right> =
D k^{-y} \delta ({\bf k}+{\bf k'})\delta (\omega+\omega')$. 
$\hat F_{\bf k}(\omega)$ denotes the Fourier transform of $F({\bf x},t)$, 
${\bf k}=(k_x,k_y)$, and $k=|{\bf k}|$. 
The process is then white in 
time but has power-law correlations in space. $y=0$ corresponds to
white-noise also in space and, in the absence of topography, it sustains the 
Kolmogorov spectrum \cite{JSC,note1}. In addition this value of $y$ has been  
observed for wind forcing on the Pacific ocean\cite{freilich85}. Thermal noise 
corresponds to $y=-4$
\cite{ll,note1}. In this case there is a fluctuation-dissipation relation 
between
noise and the viscosity term, so that the fluctuations satisfy detailed 
balance. 

To obtain the desired large-scale closed equation we have applied a 
coarse-graining procedure to the investigation of the dynamics.  
For our problem it is convenient to use the Fourier components of the 
streamfunction $\hat \psi_{{\bf k}\omega}$ or equivalently the relative 
vorticity 
$\zeta_{{\bf k}\omega}=-k^2 \hat\psi_{{\bf k}\omega}$. 
This variable satisfies:
\begin{eqnarray}
\label{fourier}
\zeta_{{\bf k}\omega} &=& G^{0}_{{\bf k}\omega} F_{{\bf k}\omega} +  
                                            \nonumber \\
&&\lambda G^{0}_{{\bf k}\omega} 
\sum_{{\bf p},{\bf q},\Omega,\Omega'} A_{{\bf k}{\bf p}{\bf q}} 
\left(   \zeta_{{\bf p}\Omega} \zeta_{{\bf q}\Omega'} + 
        \zeta_{{\bf p}\Omega} h_{\bf q}                  \right)\ \ ,
\end{eqnarray}
where the interaction coefficient is:
\begin{equation}
\label{interaction}
A_{{\bf k}{\bf p}{\bf q}}= 
(p_x q_y - p_y q_x) p^{-2} \delta_{{\bf k},{\bf p}+{\bf q}}\ \ , 
\end{equation}
the bare propagator is:
\begin{equation}
\label{propagator} 
G^{0}_{{\bf k}\omega}=(-i\omega + \nu k^2)^{-1}\ \ , 
\end{equation}
and the sum is restricted by ${\bf k}={\bf p}+{\bf q}$ and 
$\omega=\Omega+\Omega'$. ${\bf p}=(p_x,p_y)$, $p=|{\bf p}|$, and similar  
expressions hold for ${\bf q}$. 
$0<k<k_0$, with $k_0$ an upper cut-off. Following the method in Ref. 
\cite{turbulence2d}, one can eliminate the modes $\zeta^{>}_k$ with $k$ in 
the shell $k_0 e^{-\delta}< k < k_0 $ and substitute their expressions into 
the equations for the remaining low-wavenumber modes $\zeta^{<}$ with 
$0 < k < k_0 e^{-\delta}$. 
To second order in $\lambda$, the resulting equation of 
motion for the modes $\zeta^{<}$ is:
\begin{equation}
\label{result}
{\partial \nabla^2 \psi^{<} \over \partial t} + \lambda
\left[ \psi^{<} , \nabla^2 \psi^{<} +h^{<} \right] =
\nu' \nabla^4 (\psi^{<} - g h^{<}) + F' \ , 
\end{equation}
where 
\begin{equation}
\label{nu}
\nu'= \nu \left(
1 - {\lambda^2 S_2 D (2 + y ) \delta \over 32 (2 \pi)^2 \nu^3} \right),
\end{equation}
\begin{equation}
\label{g}
g(\lambda, D, \delta, \nu, y)= 
{\lambda^2 D S_2 (y+4) \delta \over 16 (2 \pi)^2 \nu^3}\ \ .
\end{equation}

$F'({\bf x},t)$ is an effective 
noise which turns out to be also a Gaussian process with mean value and 
correlations given by: 
\begin{equation}
\label{average}
<F'({\bf x},t)>=-{ \lambda^2 D S_2 (4+ y )\delta  \over 16 (2 \pi)^2 \nu^2} 
\nabla^4 h^{<},
\end{equation}
\begin{eqnarray}
\label{correlations}
\left< 
      \left(  \hat F'_k(\omega)-\left< \hat F'_k(\omega) \right>
                              \right) 
      \left(   \hat F'_{k'}(\omega')-\left< \hat F'_{k'}(\omega') \right>
                              \right)  
\right> =  \nonumber \\
D k^{-y} \delta (k+k')\delta (\omega+\omega')
\end{eqnarray} 

$S_2$ is the length of the unit circle: $2\pi$. 
Equations (\ref{result})-(\ref{correlations}) are the main
result in this Letter.  They give the dynamics of long wavelength   
modes $\psi^{<}$. They are valid for small
$\lambda$ or, when $\lambda \approx 1$, for small width $\delta$ of the 
elimination band. The effects
of the eliminated short wavelengths on these large scales are described
in the new structure of the viscosity operator and the corrections 
to the noise
term $F'$. 
The action of the dressed viscosity term $\nabla^4 (\psi^{<} - g h^{<})$ is 
no longer to drive
large scale motion towards rest, but towards a motion state 
($\approx g h^{<}$)  characterized  
by the existence of flow following the
isolevels of bottom perturbations $h^{<}$. This ground state would 
characterize the structure of the mean pattern.  
The energy in this ground state is determined by the function
$g(\lambda, D, \delta, \nu, y)$
which measures the influence of the different terms of the dynamics 
(nonlinearity, noise, viscosity). Relation 
(\ref{g}) shows that while nonlinearities \cite{note2} and noise 
increase the energy level of the ground state, high values
of the viscosity parameter would imply a reduction of the strength of the 
ground state motion due to damping effect
that viscosity exerts over small scales. 
The other mechanism that reinforces the existence of average directed motion 
comes from the fact that the dressed noise has got a mean component as a 
result of the small scale elimination.

A most interesting fact in (\ref{g}) and
(\ref{average}) is the presence of the factor $y+4$. It implies that the
tendency to form directed currents reverse sign as $y$ 
crosses the value $-4$, and that it vanishes if $y=-4$ which is the value for  
thermal noise satisfying detailed balance. The vanishing of the 
directed currents, obtained here to second order in $\lambda$, is in fact an
exact result valid to all orders in the perturbation 
expansion. This can be seen from the exact solution of the Fokker-Planck 
equation associated with Eq.(\ref{eq}) for this value of $y$\cite{longpaper}. 
This reflects the fact that noise rectification can 
not occur when detailed balance holds.   

As a consistency check we point out that if
the term representing the ambient 
vorticity $h$ is zero, classical results of two-dimensional forced 
turbulence are recovered \cite{JSC}. 
To show that the result implied by the perturbative 
expressions (\ref{result})-(\ref{correlations}) 
is really present for arbitrary $\lambda$ we check the increasing tendency 
towards average flow following the topography for increasing $\lambda$: 
Numerical simulations
of (\ref{eq}) have been conducted in a parameter regime of geophysical 
interest: 
we take $f=10^{-4} s^{-1}$ as 
appropriate for the Coriolis effect at mean latitudes on Earth 
and $\nu =200  m^2 s^{-1}$ for the viscosity, a value usual for the eddy
viscosity in ocean models. We use the numerical scheme developed in  
\cite{cummins92} on a grid of $128 \times 128$ points with a proper inclusion
of the stochastic term\cite{alvarez97}. The distance  
between grid points corresponds to 10 km, so that the total system 
size is $L=1280$ km. The amplitude of the forcing, 
$D=1\times 10^{-9}\   m^2 s^{-3}$ has  been chosen in
order to obtain final velocities of several centimeters per second. The
topographic field (shown in Fig. \ref{topography}) is randomly 
generated from a specific isotropic power 
spectrum (Fig. \ref{spectra}) with random phases. The model was run for 
$6 \times 10^5$ time steps (corresponding to 247 years) once a statistically
stationary state was reached, and some of the results for the mean
streamfunction are displayed in Figs. \ref{L01} and \ref{L1}. 
Currents with a well defined 
average sense appear. Consistently with our analytical
results, the contour levels of the mean streamfunction follow 
the topographic contours more closely the higher the value of $\lambda$ is. 
This is more quantitatively shown in Fig. \ref{correlation}  
where the linear correlation coefficient $\rho$ between the mean field 
$<\psi>$ and the underlying topography is plotted as a function of $\lambda$.  
A spectral analysis 
of the different resulting fields show that the large scales 
are better adjusted to topography, as well as the very small scales where no
significant motion is present (Fig. \ref{spectra}). 
Discrepancies are clear for the small but excited scales. This can be 
understood considering 
that the effect of viscosity on these small scales is still to drive the 
system 
towards rest. 

For negative values of $y$ noise acts more strongly on the small scales, 
where viscosity damping is more important, so that a larger noise intensity is 
needed to obtain significant large-scale directed currents. More important is 
the reversal in the sense of the currents when $y<-4$. This can be 
characterized
by the change in sign of $\rho$. For example, for $y=-6$
and $D=2\times 10^{-7}\ m^8 s^{-3}$, $\rho=-0.6$. 
More detailed results will be presented
elsewhere\cite{longpaper}. 

Concluding, the outcome of this work can be formulated as follows: 
quasigeostrophic flows develop mean patterns 
in the presence of noisy
perturbations. As relations (\ref{result})-(\ref{correlations}) show, 
the origin
of these patterns is related with nonlinearity and lack of detailed balance. 
Nonlinear terms 
couple the dynamics of small scales with the large ones and
provide a mechanism to transfer energy from the fluctuating component of
the spectrum to the mean one. This mean spectral component, that is inexistent
in purely two-dimensional turbulence \cite{thom}, 
is controlled by the shape of the
bottom boundary and characterizes the structure of the pattern.
The existence of these noise-sustained structures
has a wide range of implications
in the above mentioned fields of fluid and plasma physics. First
because it highlights the important and organizing role that noise can play  
in these systems. Secondly, it establishes the need to modify
not only the value of the parameters 
(as usually done in eddy-viscosity approaches) when performing large eddy
simulations with insufficient small-scale resolution, but also the
structure of the equations in a way determined by topography. This last
statement has been previously suggested from a heuristic point of view 
in the context of large-scale ocean models \cite{holloway92,aad}. Our results  
represent a step forward towards the justification of such approaches. 

Financial support from CICYT (AMB95-0901-C02-01-CP and MAR95-1861), DGICYT 
(PB94-1167 and PB94-1172), and from the MAST program 
MATTER MAS3-CT96-0051 (EU) is greatly acknowledged.

\newpage

\begin{figure}
\begin{center}
\def\epsfsize#1#2{0.46\textwidth}
\leavevmode
\epsffile{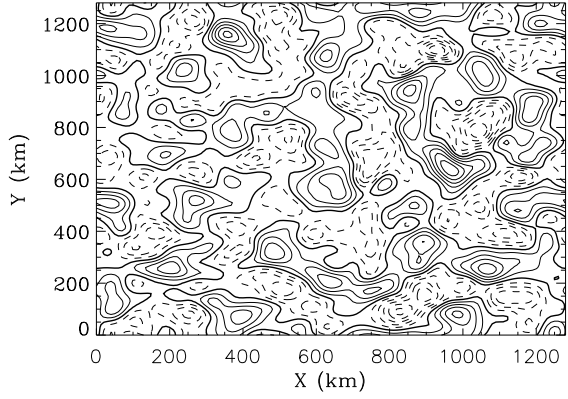}
\end{center}
\caption{\label{topography} Depth contours of a randomly generated bottom
topography. Maximum 
depth is $381.8 m$ and minimum depth $-381.8 m$ over an average depth of 
$5000 m$. 
Levels are plotted every $63.6 m$. Continuous contours are for positive 
deviations 
with respect to the mean, whereas dashed contours are for negative ones.}
\end{figure}

\begin{figure}
\begin{center}
\def\epsfsize#1#2{0.46\textwidth}
\leavevmode
\epsffile{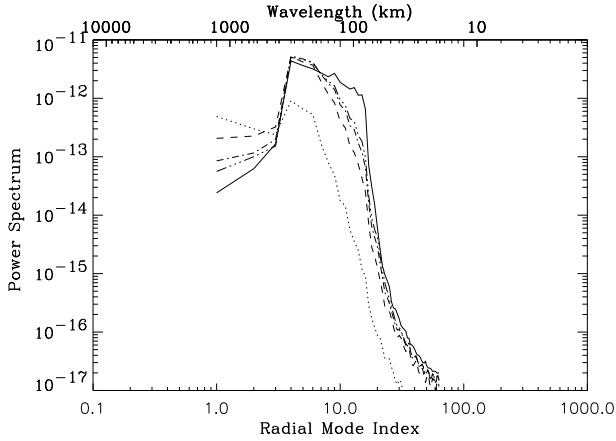}
\end{center}
\caption{\label{spectra} Comparison, as a function 
of the radial wavenumber index
 $L k^2/2\pi$, of the power spectra of  
the bottom topography (solid line) and the power spectra of
the mean streamfunction
obtained for $\lambda=0.1$ (dotted line), 0.3 (dashed line), 
0.6 (dash-dotted line) and 1 (dash-dot-dot-dot line). $y=0$ and 
$D=10^{-9} m^2 s^{-3}$. 
In order to carry out the comparison, the fields
have been normalized to have the same maximum value.
}
\end{figure}

\begin{figure}
\begin{center}
\def\epsfsize#1#2{0.46\textwidth}
\leavevmode
\epsffile{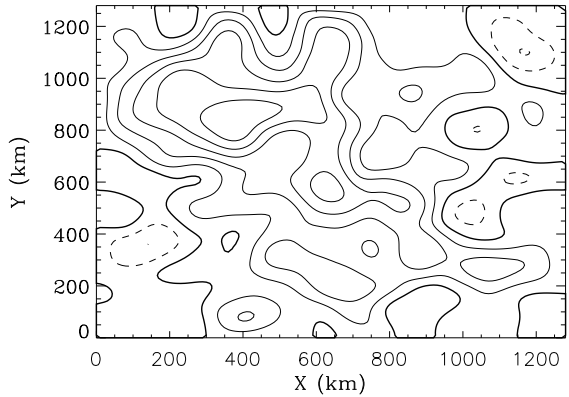}
\end{center}
\caption{\label{L01} Mean streamfunction computed by time averaging when a 
statistically stationary state has been achieved. Continuous 
contours denote positive values of the streamfunction, whereas dashed contours 
denote negative ones; $\lambda=0.1$, $y=0$, and $D=10^{-9} m^2 s^{-3}$. 
Maximum and 
minimum values are
1637.7 and -1637.7 $m^2/s$,and levels are plotted every 272.95 $m^2/s$.}
\end{figure}

\begin{figure}
\begin{center}
\def\epsfsize#1#2{0.46\textwidth}
\leavevmode
\epsffile{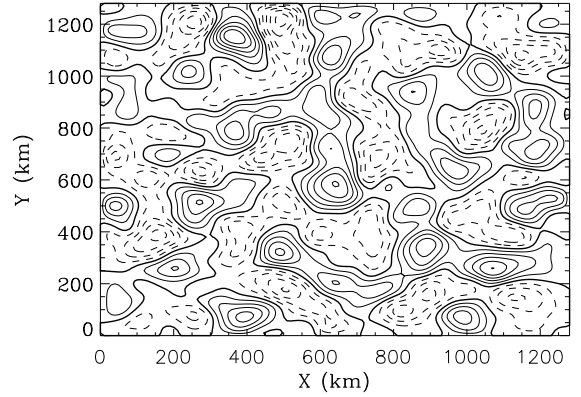}
\end{center}
\caption{\label{L1} The same as Fig. \protect{\ref{L01}} but for $\lambda=1$, 
maximum and minimum values are
991.864 and -991.864 $m^2/s$, and levels are plotted every 165.31 $m^2/s$.}
\end{figure}

\begin{figure}
\begin{center}
\def\epsfsize#1#2{0.46\textwidth}
\leavevmode
\epsffile{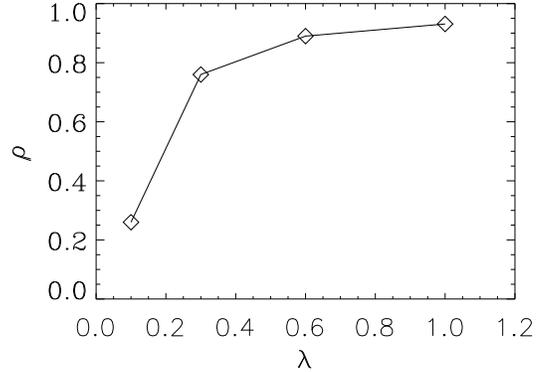}
\end{center}
\caption{\label{correlation} Linear correlation coefficient $\rho$ 
between the mean streamfunction and
topographic fields as a function of the interaction parameter $\lambda$.
$y=0$ and $D=10^{-9}m^2 s^{-3}$
} 
\end{figure}

\end{twocolumns}

\begin{references}
\bibitem{note1} It is common to work with the equation for the velocity field
instead than with the vorticity one (\protect{\ref{eq}}). 
The exponent $y'$ for the correlations 
of the noise term added to the velocity equation is related to our $y$ by 
$y=y'-2$.
\bibitem{note2} It should be noted that the parameter \protect{$\lambda$}  in
(\protect{\ref{eq}}) measures not only the size of the nonlinear term, but also
of the linear interaction between streamfunction and topography. However it is
easy to show, by using energy arguments, that the average streamfunction is 
zero if the nonlinear term is absent from (\protect{\ref{eq}}), so that the
presence  of \protect{$\lambda$} in (\protect{\ref{g}) and (\ref{average}}) 
is a
consequence of nonlinearity, not of linear scattering of the streamfunction by
the  topography.
\end{references}
\end{document}